\documentclass[aps,floatfix,nofootinbib,preprint]{revtex4}

\usepackage{graphicx}
\usepackage{epic}
\usepackage{eepic}
\usepackage{latexsym}
\usepackage{amssymb,amsmath}

\newcommand{\eq}[1]{(\ref{#1})}
\newcommand{\be}{\begin{equation}}
\newcommand{\ee}{\end{equation}}
\newcommand{\bea}{\begin{eqnarray}}
\newcommand{\eea}{\end{eqnarray}}

\newcommand{\vs}[1]{\vspace{#1 mm}}
\newcommand{\hs}[1]{\hspace{#1 mm}}

\newcommand{\vc}{\vec{k}}

\newcommand{\vv}{|0\right>}
\newcommand{\vvc}{\left<0|}

\newcommand{\hk}{\hat{k}}

\def\a{\alpha}
\def\b{\beta}
\def\cc{\gamma}
\def\C{\Gamma}
\def\d{\delta}

\def\e{\epsilon}

\def\f{\phi}
\def\fr{\frac}

\def\L{\Lambda}
\def\m{\mu}
\def\n{\nu}

\def\r{\rho}

\def\S{\Sigma}

\def\y{\eta}
\def\o{\omega}

\def\del{\partial}

\let\bm=\bibitem
\def\nn{\nonumber}

\begin{document}

\title{Vacuum arbitrariness and the Hubble tension} 
\author{Ali Kaya}
\email[]{alikaya@fas.harvard.edu}
\affiliation{\vs{3}Department of Physics, Harvard University, 17 Oxford St., Cambridge, MA 02138, USA\vs{3}\\
and\vs{3}\\
Bo\~{g}azi\c{c}i University, Department of Physics, 34342, Bebek, \.{I}stanbul, Turkey\vs{15}}

\begin{abstract}

\vs{5}

We show that the energy density of the superhorizon modes released in a non-Bunch-Davies vacuum can be arbitrarily large during inflation and it decreases like $\ln (a)/a^4$ in the subsequent radiation-dominated era. This may constitute a dark radiation component which can sufficiently alter the early cosmological evolution to alleviate the Hubble tension. 
 
\end{abstract}

\maketitle

\newpage

\section{Introduction}

There is now considerable evidence implying a genuine discrepancy between the measurements of the current Hubble parameter $H_0$ inferred from the direct local observations \cite{p0} and the cosmic microwave background (CMB) data analyzed in the concordance $\L$CDM model \cite{p1}. Although there is the possibility that the disagreement can be explained by experimental systematics (see e.g. \cite{s1} and the reply \cite{s2}), the tension clearly strengthened over time. Hence it is worth taking it seriously and to search for possible solutions potentially involving new physics. Assuming that the local observations are robust, one should obviously modify the standard $\L$CDM model to reconcile the measurements.  

The suggested resolutions in the literature either change the late-time or early-time cosmic evolution by altering the energy content of the Universe. Among the late-time resolutions one may in particular note the phantom dark energy \cite{l1}, the interacting dark energy \cite{l2} and the vacuum metamorphosis \cite{l3} models. On the other hand, the early-time modifications involve the early dark energy \cite{ede1,ede2,ede3,ede4} and the scalar field  \cite{scalar1,scalar2,scalar3} models (some specifically involving axions \cite{ax1,ax2,combo}). There are also alternative ideas like emerging spatial curvature due to nonlinear  evolution of cosmic structures \cite{curvature1}, large  impact of non-Gaussian correlations between the short and long wavelength CMB modes \cite{fluct1}, decaying dark matter \cite{ddm}, frame-dependent effective actions that mimic a cosmological constant \cite{frame} and interacting neutrinos \cite{neutrino}. It is important to note that all resolutions are tightly constrained by different observables like baryon acoustic oscillations or the CMB power spectrum peaks (see also \cite{hz1,hz2} that note a possible descending trend of $H_0$ with redshift, a new potentially constraining observable feature). 

One relatively conservative idea is to imagine the presence of early dark radiation (see e.g. \cite{p0} and \cite{dr1}), which can be quantified by the effective number of relativistic species $N_{eff}$ which normally equals 3 for the standard model neutrinos. Indeed, as noted in \cite{dr1}, for one additional species corresponding to $N_{eff}=4$, $H_0$ inferred from the Planck data increases about $7\,km/sMpc$ since the size of the sound horizon at decoupling decreases by a few percent. Such a modification is enough to ease the Hubble tension since the Planck data do not exclude this range  \cite{dr1} and the existence of dark radiation is not disfavored based on the Bayesian evidence \cite{dr2} (see also \cite{dr3} which considers an effective dark radiation component in a novel unitary gravity theory). 

\begin{figure}
	\centerline{\includegraphics[width=4cm]{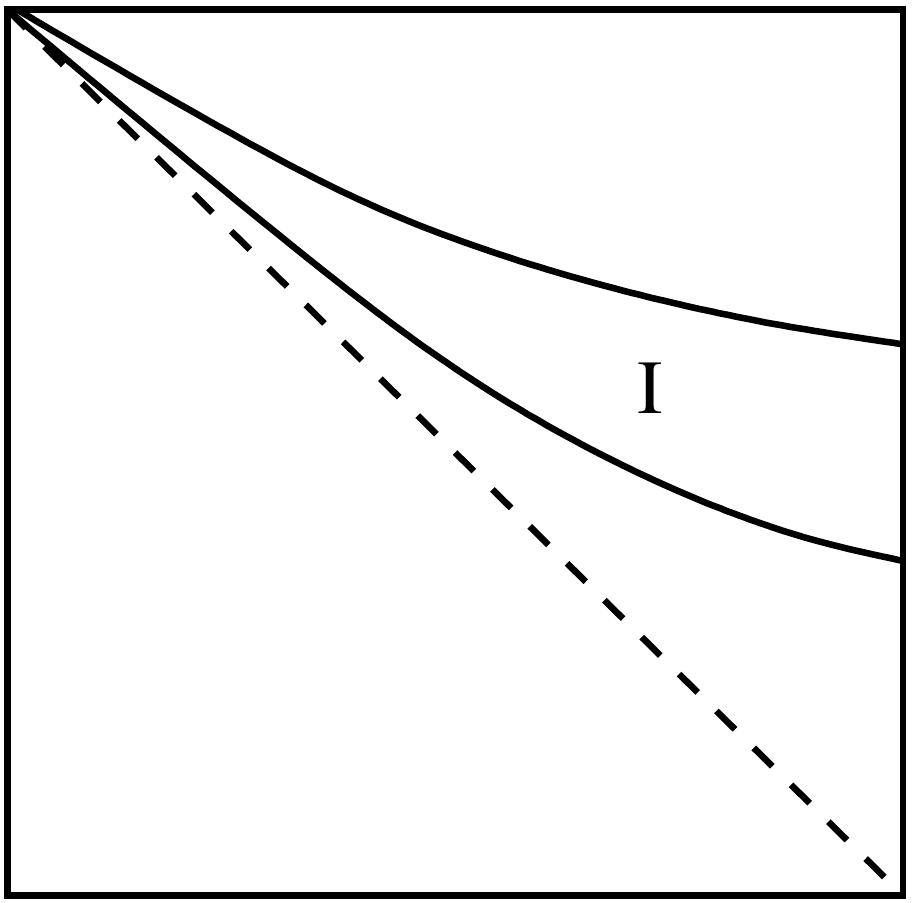}}
	\caption{The Penrose diagram of de Sitter space. Inflation corresponds to region I whose past and future should be replaced by appropriate geometries depending on the cosmological model.} 
	\label{fig1}
\end{figure}

In this work, we suggest a plausible dark radiation candidate corresponding to the superhorizon modes of a scalar field released in a {\it non-Bunch-Davies} vacuum during inflation. It is well known that there is no preferred vacuum state in a curved background \cite{fulling}, which has profound physical implications. In Minkowski spacetime the Poincare invariance selects a unique ground state, yet even demanding symmetry invariance is not enough to promote a vacuum state in curved backgrounds. The classical example is de Sitter space where there is a one complex parameter family of different vacua\footnote{Although the so-called $\a$ vacua in de Sitter space yield a peculiar analytic structure for the Green functions that seemingly become problematic in the interacting theory by producing nonlocal interactions between antipodal points \cite{i1} and loop infinities that cannot be removed by de Sitter invariant counterterms \cite{i11}, a careful treatment shows that causality can still be preserved along with renormalizability \cite{i2,i22}. Moreover, these distinctive features of the global de Sitter geometry do not directly apply to inflation which is only locally embedded in de Sitter space, see Fig. \ref{fig1}. Some implications of the $\a$ vacua for inflation were discussed, e.g. in \cite{i3,i4}.} which is invariant under the full de Sitter isometry group \cite{allen}. This arbitrariness is usually fixed by implementing some extra physical condition; the Bunch-Davies ground state is usually preferred by the argument that short-wavelength modes must mimic the flat-space propagation. However, inflation is different than the global de Sitter space (see Fig. \ref{fig1}) and one must imagine an era preceding it (indeed an incompleteness theorem \cite{bgv} requires this). In that case it is more natural to select the Bunch-Davies vacuum in the earlier period and the corresponding mode functions would carry both the positive and the negative frequencies during inflation due to Bogoliubov mixing. Therefore, there seems to be no easy way of solving the vacuum arbitrariness problem and one can equally motivate non-Bunch-Davies states for fields in inflation. 

In this paper, we focus on the modes which become superhorizon during inflation. These can be thought to represent {\it real}, rather than {\it virtual}, fluctuations. As we will see, the corresponding vacuum expectation value of the energy-momentum tensor is IR finite due to its mass dimension. While there is no UV divergence either as we focus only on the superhorizon modes, there is a trans-Planckian issue \cite{tr}; namely some of the late superhorizon modes, when followed back in time during inflation, enter the trans-Planckian regime. This shows that one should still be careful about possible renormalization (or early-time back-reaction) effects. While we will elaborate on this problem (see the Appendix), it is not the main concern of this paper, i.e. we will assume that after taking into account all renormalization (and possible backreaction) effects, one still finds an inflationary period having a Hubble scale $H$ with relatively small corrections imposed by the superhorizon energy density. 

On dimensional grounds, the energy density of massless modes during inflation must be proportional to $H^4$. This is negligible compared to the background energy density $3H^2M_p^2$ and thus cannot influence the cosmic evolution (assuming, of course, $H\ll M_p$ which is necessary for the validity of the classical gravity description). Nevertheless, we will see that in a non-Bunch-Davies vacuum characterized by the free complex parameter $\a$, the energy density becomes proportional to $|\a|^2 H^4$. If $|\a|\sim M_p/H$, this becomes comparable to the background energy density and modifies the cosmic evolution. Note that depending on the model, $|\a|$ may not be extremely large.  

Our main aim here is to find the subsequent evolution of the energy density of the $\a$-vacuum superhorizon modes after inflation. From their energy-momentum expressions, one may naively deduce that while the superhorizon modes can be characterized by the equation-of-state parameter $\o=-1/3$, the modes that reenter the horizon start oscillating and the corresponding equation-of-state parameter must be effectively $\o=1/3$. Although the falloff of the superhorizon energy density seems much slower, modes continuously cross the horizon and subhorizon modes are relatively more energetic. As a result, oscillating subhorizon modes dominate the energy density but there appears a logarithmic correction which yield an effective decay $\ln(a)/a^4$, as we will demonstrate. Since this behavior is very close to that of radiation, one can conveniently describe its impact on the cosmic evolution in terms of $N_{eff}$. We will determine the value of $\a$ that would correspond to $N_{eff}=4$, a value which is sufficient to alleviate the Hubble tension. We will also point out the impact of the $\a$ vacuum on other cosmological observables like the power spectrum.    

\section{The Vacua of the de Sitter Space}

In this section we would like to elaborate on the de Sitter symmetry group, $SO(1,4)$, invariant $\a$-vacuum found in \cite{allen}. The derivation given in \cite{allen} uses the global de Sitter geometry and the analytic properties of the Green functions. Here we utilize a direct derivation based on the mode functions, which is more suitable for inflation. 

Let $\f$ be a real massless scalar field propagating on a curved spacetime whose dynamics is governed by the standard minimally coupled action:
\be\label{a1}
S=-\fr12 \int d^4 x \sqrt{-g}\left( \nabla_\m\f\nabla^\m\f\right).
\ee
It is easy to show that this action is invariant under 
\be \label{v1}
\d \f=k^\m\del_\m\f,
\ee
where $k^\m$ is a {\it Killing} vector. We should emphasize that the background metric is taken to be fixed (and nondynamical), and hence the symmetry \eq{v1} is global and should not be confused with diffeomorphism invariance. The corresponding Noether current can be calculated as 
\be
j^\m=T^\m{}_\n\,k^\n,
\ee
where $T_{\m\n}=-2\sqrt{-g}\,\d S/\d g^{\m\n}$ is the energy-momentum tensor. The current is (covariantly) conserved $\nabla_\m j^\m=0$ and the charge 
\be\label{q} 
Q=\int_\S j^\m n_\m\, ,
\ee
which is defined as the integral over a space-like foliation with unit normal $n^\m$, is conserved. Therefore, each symmetry of the background metric yields a conserved charge which actually generates the same symmetry in the scalar field theory. Indeed, introducing the Arnowitt-Deser-Misner metric
\be
ds^2=-N^2dt^2+h_{ij}(dx^i+N^idt)(dx^j+N^jdt),
\ee
the charge can be found as 
\be
Q=-\int d^3 x\left[k^0{\cal H}+k^iP_\f\del_i\f\right],
\ee
where $k^\m=(k^0,k^i)$, ${\cal H}$ is the Hamiltonian density  
\be
{\cal H}=\fr12 \fr{N}{\sqrt{h}}P_\f^2+P_\f N^i\del_i\f+\fr12 N\sqrt{h}h^{ij}\del_i\f\del_j\f
\ee
and $P_\f$ is the conjugate momentum. Not surprisingly, one finds that (after a Legendre transformation to the Lagrangian formalism) 
\be
\{Q,\f\}=k^0\dot{\f}+k^i\del_i\f.
\ee
In the quantum theory, one further demands the invariance of the vacuum by imposing $Q\left.\vv=0$, which is a nontrivial condition as we will see below.   

We now take now a general Friedmann-Robertson-Walker background
\be
ds^2=-dt^2+a(t)^2d\vec{x}^2.
\ee
To quantize the scalar one imposes the canonical commutation relation $[\f(t,\vec{x}),P_\f(t,\vec{y})]=i\d^3(\vec{x}-\vec{y})$, where 
$P_\f=a^3\dot{\f}$. Introducing the mode decomposition 
\be
\f=\int \fr{d^3
	k}{(2\pi)^{3/2}}\left[\f_k\,e^{i\vec{k}.\vec{x}}\,a_{\vec{k}}
+\f_k^*\,e^{-i\vec{k}.\vec{x}}\,a_{\vec{k}}^\dagger\right], 
\ee
the canonical commutation relation can be satisfied by imposing $[a_{\vc},a^\dagger_{\vc'}]=\d^3(\vc-\vc')$ and 
\be\label{w}
\f_k\dot{\f}_k^*-\f_k^*\dot{\f}_k=\fr{i}{a^3}.
\ee
The mode functions obey
\be
\ddot{\f}_k+3H\dot{\f}_k+\fr{k^2}{a^2}\f_k=0\label{f}
\ee
and the vacuum is defined as
\be
a_{\vec{k}}\left.\vv=0.
\ee
Note that this is as far as one can proceed with the canonical quantization. The vacuum is associated with the mode functions, which  obey \eq{w} and \eq{f}, and otherwise are completely free. There is no preferred ground state implied by quantization. 

Let us now consider de Sitter space 
\be\label{ds}
ds^2=-dt^2+e^{2Ht}\,d\vec{x}^2.
\ee
The Bunch-Davies (BD) mode function is given by 
\be\label{bdmsl}
f_{BD}=\fr{1}{a\sqrt{2k}}\exp[ik/(Ha)]\left[1+\fr{iHa}{k}\right],
\ee
and the general solution of \eq{f} obeying \eq{w} can be written as
\be\label{mode}
\f_k=\a f_{BD}+\b f_{BD}^*,
\ee
where
\be\label{a17}
|\a|^2-|\b|^2=1.
\ee
In general both $\a$ and $\b$ can depend on $\vc$.

In addition to the obvious symmetries related to the spatial translations and rotations, there are four more Killing vectors of \eq{ds}
\bea
&&k=\fr{1}{H}\fr{\del}{\del t}-x^i\fr{\del}{\del x^i},\nn\\
&&k_{(i)}=-x^i\fr{\del}{\del t}+\left[Hx^ix^j+\fr{\d^{ij}}{2H}\left(1-H^2x^kx^k+e^{-2Ht}\right)\right]\fr{\del}{\del x^j},\label{kl}
\eea
which correspond to dilatation and special conformal transformations, respectively. One may demand the vacuum to be invariant under these symmetries, i.e. one imposes  $Q\left.\vv=0$ for each Killing vector. A straightforward calculation shows that for any mode function given in \eq{mode}, where $\a$ and $\b$ are {\it $k$-independent} constants
\be
\del_k\a=\del_k\b=0,
\ee
$Q\left.\vv=0$ is satisfied for the Killing vectors of de Sitter space. For instance, for the dilatation Killing vector given in \eq{kl} the charge operator takes the form\footnote{In putting $Q$ in this form one should apply integration by parts to remove $k$ derivatives acting on the creation/annihilation operators like $\int d^3k\, f(k)\, \del_k a_k=-\int d^3k\, \del_kf(k)\,a_k$. The surface terms vanish provided one introduces proper $i\e$ factors into the mode functions.}
\be\label{qdil}
Q=\int d^3k\left[A_k\, a_{\vc} a^\dagger_{\vc}+A_k^*\, a^\dagger_{\vc} a_{\vc}+B_k\, a_{-\vc} a_{\vc}+B_k^*\, a^\dagger_{-\vc} a^\dagger_{\vc}\right],
\ee
where 
\be
B_k=\fr{a^3}{2H}\left[\dot{\f}_k^2+\fr{k^2}{a^2}\f_k^2+3H\f_k\dot{\f}_k+Hk\dot{\f}_k\fr{d\f_k}{dk}- Hk\f_k\fr{d\dot{\f}_k}{dk}\right].
\ee
The first three terms in \eq{qdil} directly annihilate the vacuum (after normal ordering) and for invariance the last term must vanish identically. One may check that $B_k=0$ for the mode \eq{mode} if $\a$ and $\b$ are $k$-independent constants. Thus in that case the vacuum associated with \eq{mode} is invariant under the full de Sitter group. Since the overall phase of the mode is redundant and \eq{a17} must be obeyed, there remains a one-parameter family of vacua parametrized by a completely free complex number $\a$, as first shown in \cite{allen}. 

Although there are concerns about the validity of the $\a$ vacuum in the interacting theory \cite{i1}, these are mostly related to the analytical structure of the Green functions in the global de Sitter space (and they do not necessarily imply any physically unacceptable property \cite{i2}). In the case of inflation, one only considers nearly exponential expansion in a finite duration and the in-in (Keldysh-Schwinger) perturbation theory must be safe provided that the $i\e$ prescription is properly applied. Thus, there is no reason to doubt about the validity of the perturbative interacting theory in the $\a$ vacuum. 

\section{The Energy Density of the Superhorizon Modes}

Our aim in this section is to determine the subsequent evolution of the energy density of the modes that become {\it superhorizon at the end of inflation.} As pointed out in the Introduction, these modes can be thought to represent real quantum fluctuations rather than being virtual. Therefore, it is physically viable to assume that their energy density contributes to the spacetime curvature in general relativity. The scale factor of the universe describing an inflationary era which is immediately followed by a radiation epoch can be written as 
\be\label{inrad}
a(t)=\begin{cases}a_I\, e^{Ht}\hs{15}t_i\leq t\leq t_I \vs{2}\cr
\left(t/t_0\right)^{1/2}\hs{10}t_I\leq t,
\end{cases}
\ee
where $a_I$ is a constant. Matching the scale factor at $t_I$ gives
\be
a_I=e^{-1/2}\sqrt{H_0/H},\hs{10}t_I=1/2H,\label{23}
\ee
where $H_0=1/2t_0$ is the Hubble parameter at $t_0$. We note that $t_0$ can be any time of interest in the radiation era. 

It is useful to introduce the conformal time $d\y=dt/a$ and shift it so that at the inflation-radiation interface it is equal to
\be
\y_I=-\fr{1}{\sqrt{HH_o}}.
\ee
Then, the scale factor in terms of the conformal time is given by 
\be\label{mct}
a(\y)=\begin{cases}-\fr{1}{H\y}\hs{15}\y_i\leq \y\leq \y_I, \vs{2}\cr
\fr{\y-2\y_I}{2t_0}\hs{14}\y_I\leq \y,
\end{cases}
\ee
where the continuity at $\y=\y_I$ can be verified easily. Defining 
\be
\m_k=a\f_k
\ee
the mode equation \eq{f} implies 
\be\label{mueta}
\m_k''+\left(k^2-\fr{a''}{a}\right)\m_k=0,
\ee
where a prime denotes a derivative with respect to $\y$. During radiation $a''=0$, and therefore the two solutions of \eq{mueta} are $\cos(k\y)$ and $\sin(k\y)$. 

Consider a moment $\y$ during the radiation epoch. A given comoving scale can be either subhorizon/superhorizon in its entire history or it reenters the horizon by the time $\y$. This behavior can be summarized as follows: 
\bea
&& 0<k< \fr{1}{\y+|\y_I|}:\hs{10}\textrm{always superhorizon,}\nn\\
&& \fr{1}{\y+|\y_I|}<k< \fr{1}{|\y_I|}:\hs{5}\textrm{superhorizon}\to\textrm{subhorizon},\label{ssh}\\
&& \fr{1}{|\y_I|}<k:\hs{25}\textrm{always subhorizon.}\nn
\eea
The modes of our interest that become superhorizon at the end of inflation are given by
\be \label{moi}
0<k< \fr{1}{|\y_I|}.
\ee
Some of these modes become subhorizon later on as pointed out in \eq{ssh}. From the energy-momentum tensor $T_{\m\n}=\nabla_\m\phi \nabla_\n\phi-\fr{1}{2}g_{\m\n}(\nabla\phi)^2$, one can  calculate the following vacuum-expectation values for these modes
\bea
&&\r=\frac{1}{4\pi^2 a^4}\int_0^{1/|\y_I|} k^2\left[
|\dot{\f}_k|^2+\fr{k^2}{a^2}|\f_k|^2\,\right]dk,\label{se} \\ 
&&P=\frac{1}{4\pi^2 a^4}\int_0^{1/|\y_I|} k^2\left[ |\dot{\f}_k|^2-\fr{k^2}{3a^2}|\f_k|^2\,\right]dk.\nn
\eea
Since the momentum integral is cut off at a fixed comoving scale, the energy conservation $\dot{\r}+3H(\r+P)=0$ is obeyed provided that the mode equation \eq{f} holds. 

We will assume that the modes are released in the $\a$ vacuum so that 
\be\label{av}
\m_k=\fr{\a}{\sqrt{2k}}\left[1-\fr{i}{k\y}\right]e^{-ik\y}+\fr{\b}{\sqrt{2k}}\left[1+\fr{i}{k\y}\right]e^{ik\y}\, ,\hs{3}\y<\y_I,
\ee
during inflation. When the modes evolve through the radiation era, their solution takes the form
\be\label{mrad}
\m_k=\m_k^I\cos\left[k(\y-\y_I)\right]+\fr{\m_k^I{}'}{k}\sin\left[k(\y-\y_I)\right], \hs{3}\y_I<\y,
\ee
where $\m_k^I=\m_k(\y_I)$ and $\m_k^I{}'=\m_k'(\y_I)$ must be read from \eq{av} to match the mode functions at $\y=\y_I$. The Bunch-Davies ground state corresponds to $\a=1$, $\b=0$ and it is motivated by having a single (negative) frequency solution which asymptotes to the flat-space mode function at short wavelengths when $k\y\to\infty$. However, a pure positive or negative frequency mode picks up both frequencies when it passes from one cosmological epoch to another (see e.g. \eq{mrad} for $\a=1$). It is natural to assume that the modes were born before inflation as suggested by the incompleteness theorem of \cite{bgv}, and hence it is also natural to take the general mode solution carrying both the positive and the negative frequencies during inflation. (As discussed in the previous section, demanding the symmetry invariance does not help either in preferring a vacuum state in this setup.) Without a detailed model that describes the cosmology before inflation, one cannot determine $\a$; therefore, at this time it should be treated as a free parameter to be determined by observations. 

We can now calculate the energy density of the modes by using \eq{av} during inflation or \eq{mrad} during the radiation era. Using \eq{av} in \eq{se}, one can see that the energy density stored in the modes \eq{moi} {\it at the end of inflation} becomes 
\be\label{eei}
\r(\y_I)=\fr{|\a|^2+|\b|^2+Re\, \{\a\b^*z\}}{8\pi^2}\,H^4,
\ee
where $z\simeq -1.4+0.8 i$. Not surprisingly the energy density is proportional to $H^4$ as can be expected by dimensional analysis. Still, it is also proportional to $|\a|^2$ and in principle one can have $\r(\y_I)\gg H^4$ if $|\a\|\gg1$. This possibility helps one to address the Hubble tension as we will show in the next section. 

Before discussing how this energy density evolves after inflation in the radiation era, one may note its earlier behavior. The physical wave number of the mode which becomes superhorizon at the end of inflation is  $k/a(\y_I)=H$ and this enlarges to $e^N H$ when followed back in time with $N$ e-folds. The modes clearly enter in the trans-Plankian regime at an exponentially increasing rate. Using \eq{se} at a generic time at inflation gives 
\be\label{ris0}
\r(\y)=\left[(|\a|^2+|\b|^2)\fr{\y^2(\y^2+\y_I^2)}{\y_I^4}+Re\left\{ 3\a\b^* +\a\b^* e^{-2i\y/\y_I}\left[\fr{4\y^2}{\y_I^2}-i\fr{6\y}{\y_I}-3\right]
\right\}
\right]\fr{H^4}{16\pi^2}.
\ee
One may check that this $\r(\y_I)$ agrees with \eq{eei}. On the other hand, when $\y/\y_I=e^N$ corresponding an earlier time with $N$ e-folds to the end of inflation, the dominant contribution $e^{4N}H^4$ exceeds the background energy density when $N$ is sufficiently large. Therefore the backreaction of these modes cannot be neglected and \eq{ris0} loses its validity at earlier times corresponding to $|\y|\gg|\y_I|$. 

The trans-Planckian problem can be addressed in different ways and obviously renormalization of the physical quantities should be taken into account (in the Appendix we discuss possible regularization schemes). We avoid this discussion by simply focusing on the superhorizon modes at a given time $\y$ in inflation, which corresponds to the range $0<k< 1/|\y|$ as opposed to \eq{moi}. In that case the upper limit of the integrals in \eq{se} must be replaced by $1/|\y|$ which gives  
\bea
&&\r(\y)=\fr{|\a|^2+|\b|^2+Re\, \{\a\b^*z_1\}}{8\pi^2}\,H^4,\label{ris}\label{35}\\
&&P(\y)=\fr{Re\, \{\a\b^*z_2\}}{48\pi^2}\,H^4,\nn
\eea
where $z_1=z\simeq -1.4+0.8 i$ and $z_2\simeq 3.0-2.2i$. One should be careful here since the energy conservation is not satisfied as the comoving integration range is chosen to be time dependent i.e. extra modes keep entering the superhorizon regime continuously. Indeed one can verify that $\dot{\r}+3H(\r+P)\not=0$ and the nonzero right-hand side equals the influx of the modes which is given by $(-1/\eta)'\times$ [the energy density per comoving wave number at $k=-1/\y$]. As long as backreaction is neglected and \eq{ris} is not used as a source for gravity, energy nonconservation  is not a problem since \eq{ris} simply gives the energy density of an {\it open system.} Note that $\r(\y_I)$ calculated from \eq{ris} is equal to \eq{eei}. It is thus reasonable to assume that the energy density of the superhorizon modes at the end of inflation is given by \eq{eei}, which can either be obtained from \eq{ris0} or \eq{ris}. The mode momentum integrals carried out above are all IR finite since the integrands are not  singular as $k\to0$ and there is also no UV divergence since the integrals are cut off at a suitable scale.  

Let us now determine the evolution of \eq{eei} in the radiation era. This can be done exactly by using \eq{mrad} in \eq{se}, which gives 
\bea
\r(\y)=&&\int_0^{1/|\y_I|}dk\,\fr{H_0^2}{8\pi^2a^6}\left\{[k^2|\m_k^I|^2+|\m_k^I{} '|^2][1+2k^2(\y-2\y_I)^2]\right.\nn\\
&&+\left[k^2|\m_k^I|^2-|\m_k^I{}'|^2-4k^2Re\left\{\m_k^I{}^*\m_k^I{} '\right\}(\y-2\y_I)\right]\cos(2k(\y-\y_I))\nn\\
&&+k\left.\left[2Re\left\{\m_k^I{}^*\m_k^I{}'\right\}+2[k^2|\m_k^I|^2-|\m_k^I{}'|^2](\y-2\y_I)\right]\sin(2k(\y-\y_I))\right\}. \label{36}
\eea
Recall that  $\m_k^I=\m_k(\y_I)$ and $\m_k^I{}'=\m_k'(\y_I)$ where $\m_k$ is given in \eq{av}. It is possible to evaluate this integral exactly; however the final result is cumbersome and not very illuminating since there are terms with many different structures. We  note that as $k\to0$ the integrand in \eq{36} becomes
\be
\fr{|\a|^2+|\b|^2-2 Re\{\a\b^*\}}{8\pi^2a^2}\,H^2\,k+{\cal O}(k^0),
\ee
thus the integral is IR finite, which is not surprising since the mass dimension of the integrand is 3. To simplify the computation, one may note that in the range of $k$ integration $[0,1/|\y_I$], the exponential phase factor $e^{\pm ik\y_I}$ appearing in  $\m_k^I$ and $\m_k^I{}'$ does not actually oscillate since the phase $k\y_I$ takes values in the interval $[0,1]$ in radians. Therefore, to an excellent approximation, $e^{\pm ik\y_I}$ can be replaced by its power series up to a desired order. Keeping the second-order terms in this power series and integrating \eq{36} yields 12 different terms. At late times the energy density can be seen to take the following form
\be\label{rlt}
\r(\y)\simeq\fr{|\a-\b|^2}{8\pi^2}\fr{H^2}{H_0^2}\fr{1}{\y^4}\ln\left(\fr{\y}{|\y_I|}\right)+{\cal O}\left(\fr{H^2}{H_0^2\y^4}\right)+
{\cal O}\left(\fr{H^2}{H_0^2\y^4}\fr{\y_I}{\y}\right), \hs{5}\y\gg|\y_I|. 
\ee 
The leading-order contribution containing the logarithm is proportional to $|\a-\b|^2$. To understand this dependence, one may note that in the deep superhorizon regime \eq{av} evaluated at $\y=\y_I$ can be approximated as  
\be\label{m39}
\m_k^I\simeq\fr{-i(\a-\b)}{\sqrt{2}\,\y_I\,k^{3/2}},\hs{5}\m_k^I{}'\simeq\fr{i(\a-\b)}{\sqrt{2}\,\y_I^2\,k^{3/2}}.
\ee
It turns out that using this approximate form in \eq{36} also yields the same leading-order term in \eq{rlt}, which explains the dependence on $\a-\b$. This is a curious  example of an IR logarithm which shows up in an epoch following inflation.

By noting 
\bea
&&a(\y)\simeq \y H_0,\hs{5}\y\gg|\y_I |\nn\\
&&a(\y_I)=|\y_I|H_0=\sqrt{\fr{H_0}{H}},
\eea
and using \eq{eei}, the evolution of the energy density can be written as 
\be\label{revson}
\r(\y)\simeq \ln\left(\fr{a(\y)}{a(\y_I)}\right)\fr{a(\y_I)^4}{a(\y)^4}\,\r(\y_I). 
\ee
Here we assume that the phases of $\a$ and $\b$ are generically chosen so that $|\a-\b|$ has the same order of magnitude as $|\a|$. If this is not satisfied and $|\a-\b|\ll|\a|$, one must look at the second-order term in \eq{rlt} which corresponds to the falloff $1/a^4$. In any case, we see that the energy density stored in the modes which become superhorizon at the end of inflation decreases very similarly to that of radiation. 

\section{Implications for the Hubble Tension} 

In this section we will discuss how the above dark radiation component can help to resolve the Hubble tension. From \eq{eei} the initial energy density of the dark radiation at the end of inflation can be taken as
\be\label{apdr}
\r_D\simeq\fr{|\a|^2}{4\pi^2}\,H^4.
\ee
We approximate its time evolution by the simple power-law decrease $1/a^4$ by neglecting the slow change caused by the logarithm. The energy density in the known radiation component can be parametrized as
\be\label{rvs}
\r_R=g_*\fr{\pi^2}{30}T^4,
\ee
where $T$ is the equilibrium temperature and $g_*$ is the effective number of relativistic degrees of freedom given by 
\be
g_*=g_B\,\left(\fr{T_B}{T}\right)^4+g_F\,\fr78 \left(\fr{T_F}{T}\right)^4.
\ee
Here $g_B$ and $g_F$ are the bosonic and the fermionic degrees of freedom with temperatures $T_B$ and $T_F$, respectively, and the factor $7/8$ accounts for the difference in the Fermi/Bose statistics. When $T\gg m_e$, $g_*=11/2$ which corresponds to $g_B=2$ for photons and $g_F=4$ for $e^\pm$ pairs, all having the same temperature $T$. After $e^\pm$ annihilation, the photon temperature increases since the entropy  $g_*(aT)^3$ remains constant. After decoupling, $g_*=2$ and thus the photon temperature increases by the factor $(11/4)^{1/3}$ relative to other relativistic species like neutrinos. As a result the known component of the radiation energy density can be parametrized like  
\be\label{totr}
\r_R=\left[1+\fr78 \left(\fr{4}{11}\right)^{4/3}\, N_{eff}\right]\r_\cc,
\ee
where $\r_\cc$ is the photon energy density and $N_{eff}=3$ for the three standard model neutrino species. 

As pointed out in the Introduction, a new neutrino species which would give $N_{eff}=4$ can resolve the Hubble tension without violating the observational constraints. Assuming that the background inflaton energy density after inflation is converted by reheating to known radiation with $N_{eff}=3$, one obtains
\be\label{rcc}
\left[1+\fr{21}{8} \left(\fr{4}{11}\right)^{4/3}\right]\r_\cc=3H^2M_p^2
\ee 
Then, to get a  total radiation energy density with $N_{eff}=4$ one must have
\be
3H^2M_p^2+\r_D=\left[1+\fr{28}{8} \left(\fr{4}{11}\right)^{4/3}\right]\r_\cc,
\ee
which implies 
\be
\r_D=\fr{7}{8} \left(\fr{4}{11}\right)^{4/3}\r_\cc.
\ee
From \eq{apdr} and \eq{rcc} this yields
\be\label{est}
|\a|\simeq 4 \,\fr{M_p}{H}.
\ee
Typically, the inflationary Hubble scale is taken as $H=10^{-3}M_p$, which gives $|\a|\simeq 4000$. Fortunately this is not an extremely large number but its naturalness cannot be judged either. Remember that we have motivated non-Bunch-Davies vacuum by referring to an epoch preceding inflation and without having the complete history it is difficult to estimate the magnitude of $\a$.  

Consistency requires that the same (or very similar) vacuum choices must be made for all quantum fields in a model. This is particularly necessary if $\a$ vacuum is imagined to arise because of mode crossing towards inflation as we have argued. In that case, basic cosmological observables should also be modified accordingly. The power spectrum is the most crucial of such observables and one must in particular ensure that the scale-freeness is not spoiled. Fortunately, from the $\a$-vacuum mode function \eq{av} one still gets a scale-free power spectrum 
\be
P_k\equiv |\f_k|^2\simeq |\a-\b|^2\,\fr{H^2}{2k^3},\hs{5}k\y\ll1.
\ee
Compared to the Bunch-Davies vacuum, only the amplitude is modified by the constant factor $|\a-\b|^2$. Therefore, neither the scalar-to-tensor ratio nor the running of the index are altered in the $\a$ vacuum.  

Finally in models containing more than one scalar field, the estimate of $|\a|$ in \eq{est} is reduced by the number of fields since each contributes to the energy density the same amount. Moreover, the neglected slow logarithmic growth also helps to reduce the value of $|\a|$. In this work we treated the dark radiation as a noninteracting component which is decoupled from the thermal plasma of the standard model particles. It would be interesting to study the other possibility, e.g. to work out the cosmological consequences of a dark radiation consisting of Higgs particles. We leave this curious problem as a topic for future research. 

\appendix* 

\section{Regularization in the $\a$-vacuum}

In this appendix, we discuss how the UV infinities of the energy-momentum tensor expectation values can be regularized in the $\a$ vacuum. Since the high-energy limit of the mode functions differs from the flat-space behavior, it is natural to expect some distinguishing features to appear. Yet, it will be quite surprising to observe the failure of the regularization procedure in a free theory given the fact that the $\a$-vacuum interacting field theory in the de Sitter space physically makes sense \cite{i2}. In any case, any unwanted behavior that might possibly arise in the regularization/renormalization process can be avoided by taking a truncated $\a$-vacuum if necessary,\footnote{We thank the anonymous referee for suggesting this possibility.} by choosing $\a$ and $\b$ as $\vc$-dependent functions. (Of course this dependence breaks some of  de Sitter symmetries, but the slow-roll inflationary models have a smaller number of symmetries anyway, so this is not an immediate point of concern.) Since for the main conclusion of this paper it is enough to take the $\a$ vacuum only for the superhorizon modes, the issues about the short-distance behavior can be evaded in this way. 

On the other hand, we have mentioned that one way of motivating the $\a$ vacuum during inflation is to presume the existence of a preceding era during which the modes are released in their corresponding Bunch-Davies ground states. It is possible to see that such a model naturally yields states that look like a truncated $\a$ vacuum during inflation, where the subhorizon excitations have the Bunch-Davies mode functions and the superhorizon excitations have the $\a$-vacuum mode functions. To see how this may arise note that both $\a$ and $\b$ become $\vc$-dependent coefficients in inflation due to the change in the modes evolving from one era to another. This change can be interpreted as a particle creation process and the effect must depend on the physical momentum $k/a$, hence, one should have $\a=\a(k/a)$ and $\b=\b(k/a)$. The spacetime curvature does not alter the short-wavelength subhorizon behavior significantly, and therefore one expects
\bea
&&\a\left(k/a\right)\simeq1,\hs{5}k/a\gg H,\nn\\
&&\b\left(k/a\right)\simeq0,\hs{5}k/a\gg H,
\eea
since the modes were originally released in their Bunch-Davies ground states. This corresponds to a suitable truncated $\a$ vacuum having the standard UV behavior. To resolve the Hubble tension it is enough to have the estimate \eq{est} for the superhorizon modes, 
\be
\left|\a(k/a)\right|\simeq 4\fr{M_p}{H},\hs{5}k/a<H. 
\ee
Note that this setup is equivalent to the resolution of the trans-Planckian problem discussed around \eq{35}. 

Yet, one may still wonder how regularization works for the pure $\a$ vacuum in de Sitter space. Let us illustrate this for the two-point function using dimensional regularization. For this calculation one must carry out the quantization procedure in $(d+1)$-dimensional de Sitter space. In that case, the massless scalar mode equation and the Wronskian condition in $d$ spatial dimensions become
\be
\ddot{\f}_k^{(d)}+d\,H\,\dot{\f}_k^{(d)}+\fr{k^2}{a^2}\f_k^{(d)}=0,\hs{10}\f_k^{(d)}\dot{\f}_k^{(d)*}-\f_k^{(d)*}\dot{\f}_k^{(d)}=\fr{i}{a^d}.\label{a1}
\ee
The de Sitter space Bunch-Davies mode function is given by   
\be
f_{BD}^{(d)}=\fr{1}{a^{d/2}}\sqrt{\fr{\pi}{4H}}\exp\left(i\pi d/2\right)H_{d/2}^{(1)}\left(\fr{k}{aH}\right),\label{a2} 
\ee
where $H_n^{(1)}$ is the Hankel function of the first kind. The $\a$-vacuum mode function is defined as usual by 
\be\label{amode}
\f_k^{(d)}=\a f_{BD}^{(d)}+\b f_{BD}^{(d)}{}^*.
\ee
We will regularize the two-point function in the coincident limit as
\be\label{a4}
\vvc \f^2\vv= \m^{\d}\int\, \fr{d^d k}{(2\pi)^d}\,|\f_k^{(d)}|^2,\hs{10}\d\equiv3-d,
\ee
where we introduced a mass scale $\m$ to compensate the dimensional mismatch of the fields as is customary in dimensional regularization. After the scaling $k=aH\hat{k}$, \eq{a4} becomes
\be
\vvc \f^2\vv=\fr{e^2H^2}{32\pi^2}\left(\fr{2\pi\m}{H}\right)^\d\int\, d^{3-\d}\hk\,\left|\a\,H_{(3-\d)/2}^{(1)}(\hk)+\b\,H_{(3-\d)/2}^{(2)}(\hk)\right|^2,\label{aa4} 
\ee
where $H_n^{(2)}$ is the Hankel function of the second kind, which obeys $H_n^{(2)}=H_n^{(1)*}$ for real $n$. For the dimensional regularization to work, the integral in \eq{aa4} must obey 
\be\label{a5}
\lim_{\d\to0}\,\int\, d^{3-\d}\hk\,\left|\a\,H_{(3-\d)/2}^{(1)}(\hk)+\b\,H_{(3-\d)/2}^{(2)}(\hk)\right|^2=\fr{F_0}{\d}+F_1,
\ee
where $F_0$ and $F_1$ are finite numerical factors. 

\begin{figure}
	\centerline{\includegraphics[width=5cm]{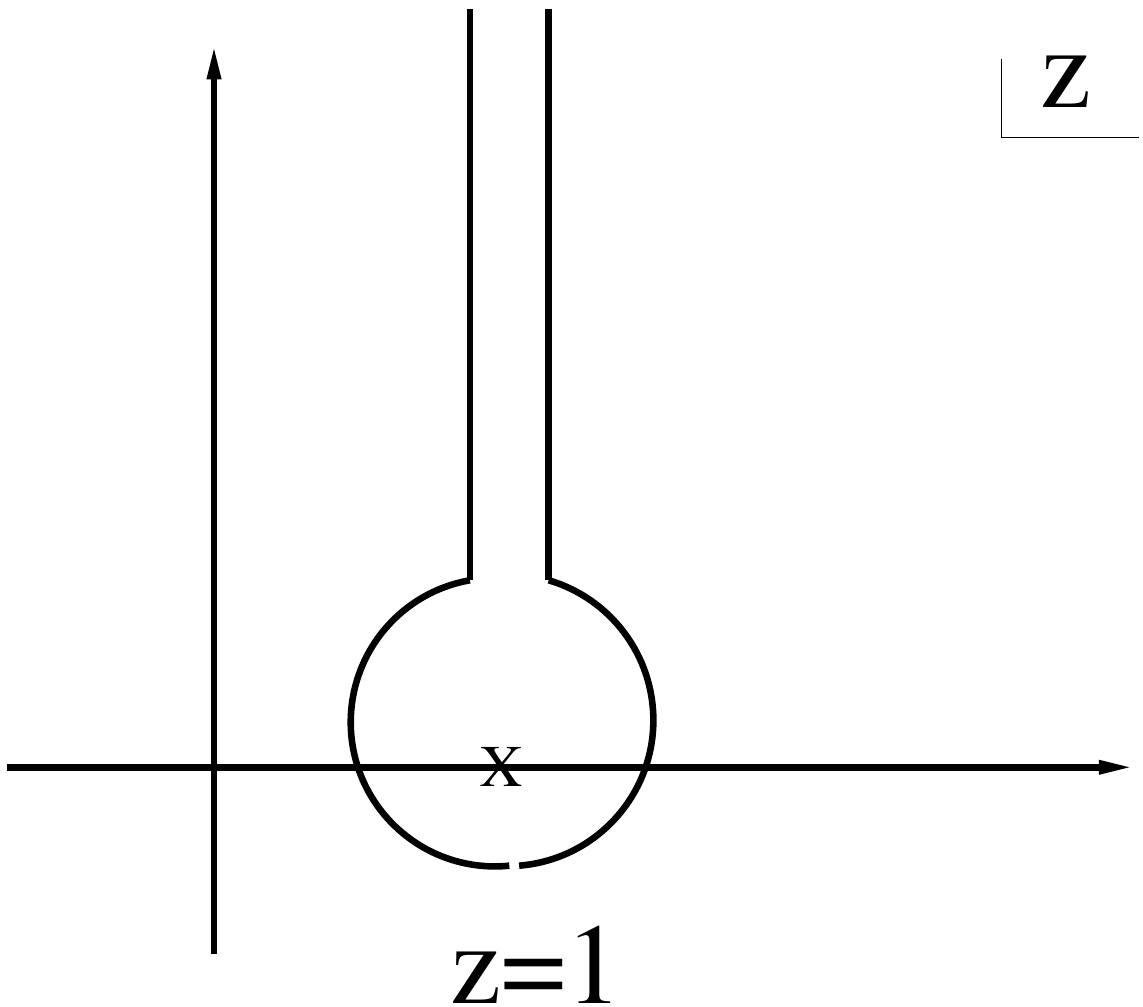}}
	\caption{The integration contour defining the Hankel function $H_n^{(1)}$.}
	\label{fig2}
\end{figure}

To see that \eq{a5} is actually satisfied, one may use the following integral representation of the Hankel function  \cite{33}
\be\label{aint}
H_n^{(1)}(\hk)=\fr{\C[\fr12-n](\fr12 \hk)^n}{\pi^{\fr32}i}\oint_{1+i\infty}^{(1^+)}e^{i\hk z}(z^2-1)^{n-\fr12}\,dz,
\ee
where the simple loop contour starts at $z=1+i\infty$ in the complex $z$ plane, circles $z=1$ once in the counter-clockwise direction and returns to $z=1+i\infty$ (see Fig. \ref{fig2}). The nonanalytic function $(z^2-1)^{n-1/2}$ is continuous on the contour, where its principal value is read from the intersection of the path with the $(1,\infty)$ line. This representation is valid for any $n\not=$ half-integer and when $\arg(\hk)<\pi/2$. For us $n=(3-\d)/2$, and one can see that the loop integral about $z=-1$ vanishes when it shrinks to the zero size. Therefore, one may assume that $z$ always has a positive imaginary piece and $\exp(i\hk z)$ has an exponentially decaying factor. In evaluating \eq{a5}, one can now use \eq{aint} for one of the Hankel functions (after expanding the absolute value squared). In that case, the momentum integral $\int d^{3-\d}\hk$ becomes well defined as $\hk\to\infty$ because of the exponential decay coming from $\exp(i\hk z)$. For example, one can rewrite 
\bea
&&\int\,d^{3-\d}\hk\,H_{(3-\d)/2}^{(1)}(\hk)H_{(3-\d)/2}^{(2)}(\hk)=\nn
\\
&&\hs{7}\fr{\C[\fr{\d}{2}-1](\fr12 \hk)^{(3-\d)/2}}{\pi^{\fr32}i}\int\, d^{3-\d}\hk\,\oint_{1+i\infty}^{(1^+)}e^{i\hk z}(z^2-1)^{1-\fr{\d}{2}}\,
H_{(3-\d)/2}^{(2)}(\hk)\,dz
\eea
where the integrals are well defined even when $\d=0$. As a result, \eq{aint} lets one to hide the original momentum integral infinity in the gamma function and in the limit $\d\to0$ \eq{a5} becomes 
\be
\lim_{\d\to0}\,\int\, d^{3-\d}\hk\,\left|\a\,H_{(3-\d)/2}^{(1)}(\hk)+\b\,H_{(3-\d)/2}^{(2)}(\hk)\right|^2=\lim_{\d\to0}\,\C\left[\fr{\d-2}{2}\right]G(\d),
\ee
where $G(\d)$ is an analytic function of $\d$ that also depends on $\a$ and $\b$. This verifies \eq{a5}, where the terms $F_0$ and $F_1$ can be read from the expansions of $G(\d)$ and the gamma function $\C[\d/2-1]$ as $\d\to0$. Consequently \eq{a4} becomes
\be
\vvc \f^2\vv=\fr{e^2H^2}{32\pi^2}\,\lim_{\d\to0}\,\left(1+\d\ln(2\pi\m/H)\right)\left(F_0/\d+F_1\right)
\ee 
where the singular $1/\d$ piece must be canceled out by a counterterm. We thus see that the $\a$ vacuum and the Bunch-Davies vacuum are similar from the dimensional regularization perspective.   

Finally, it is instructive to consider a brute-force cutoff regularization to see what differences appear between the vacua. Introducing a (comoving) momentum cutoff $\L$ gives the energy density and the pressure as 
\bea
&&\r=\frac{1}{4\pi^2 a^4}\int_0^{\L} k^2\left[
|\dot{\f}_k|^2+\fr{k^2}{a^2}|\f_k|^2\,\right]dk,\label{apse} \\ 
&&P=\frac{1}{4\pi^2 a^4}\int_0^{\L} k^2\left[ |\dot{\f}_k|^2-\fr{k^2}{3a^2}|\f_k|^2\,\right]dk.\nn
\eea
These integrals can be carried out exactly for the $\a$ vacuum, which of course have diverging terms as $\L\to\infty$. A straightforward calculation in the limit $\L\to\infty$ gives (recall that $a=-1/H\eta$)
\bea
\r=&&\fr{H^4}{16\pi^2}\,Re\left\{e^{2i\L\y}\,\a^*\b\left(4\L^2\y^2+6i\L\y-3\right)\right\}+\fr{H^4(|\a|^2+|\b|^2)}{16\pi^2}\left(\L^4\y^4+\L^2\y^2\right)\nn\\&&+\fr{3H^4(\a\b^*+\a^*\b)}{32\pi^2}+{\cal O}\left(\fr{1}{\L}\right),\nn\\
P=&&\fr{H^4}{16\pi^2}\,Re\left\{e^{2i\L\y}\,\a^*\b\left(\fr{8i}{3}
\L^3\y^3-\fr{16}{3}\L^2\y^2-6i\L\y+3\right)\right\}\nn\\&&+\fr{H^4(|\a|^2+|\b|^2)}{48\pi^2}\left(\L^4\y^4-\L^2\y^2\right)-\fr{3H^4(\a\b^*+\a^*\b)}{32\pi^2}+{\cal O}\left(\fr{1}{\L}\right).
\eea
As opposed to the dimensional regularization, in the cutoff method the $\a$ vacuum shows a peculiar behavior that involves oscillating infinities (note that these vanish for the Bunch-Davies vacuum which has $\b=0$). On the other hand, like the Bunch-Davies vacuum, the nonoscillating part of the energy-momentum tensor which has quartic infinities has the equation-of-state parameter $\o=1/3$, while the part containing quadratic infinities has the equation-of-state parameter $\o=-1/3$. Canceling these infinities requires counterterms that break the general covariance (note that this is also the case for the Bunch-Davies vacuum). This is not a surprising result, which is actually similar to the loss of Lorentz invariance in the flat-space quantum field theory when the brute-force cutoff regularization is applied. Remarkably, however, the finite $\L^0$-order terms correspond to a cosmological constant that simply modifies the bare value. Assuming that the diverging terms are removed, one can still get a de Sitter space with the $\a$ vacuum.

\acknowledgments{I am grateful for the support of the Harvard SAR Program and the Harvard Physics Department. I also thank the colleagues at the Harvard Physics Department for their hospitality.}

\end{document}